\documentclass[11pt]{article}
\usepackage{verbatim}       % comment out chunks of file
\usepackage{amsfonts}       % nice letters
\usepackage{amsmath}        % subequations, boxed formulas
\usepackage{amssymb}
\usepackage{bm}         % bold math
\usepackage{epsfig}

\def\ben{\begin{equation}}
\def\een{\end{equation}}
\def\half{{\textstyle{1\over2}}}

   \let\d=\delta \let\e=\varepsilon

\let\pa=\partial
\def\ba{\begin{array}}
\def\ea{\end{array}}

\def\dalemb#1#2{{\vbox{\hrule height .#2pt
        \hbox{\vrule width.#2pt height#1pt \kern#1pt
                \vrule width.#2pt}
        \hrule height.#2pt}}}

\newcommand{\bea}{\begin{eqnarray}}
\newcommand{\eea}{\end{eqnarray}}
\def\Tr{{\rm Tr}}

\def\be{\begin{equation}}
\def\ee{\end{equation}}
\def\bea{\begin{eqnarray}}
\def\eea{\end{eqnarray}}

\def\d{\delta}

\def\R{{\mathbb R}}
\def\pa{\partial}

\begin{document}
\begin{flushright}
NSF-KITP-07-142 \\
WIS/08/07-JUNE-DPP \\
SLAC-PUB-12570 \\
\end{flushright}

\begin{center}
\vspace{1cm} { \LARGE {\bf A phase transition in commuting}}
\vspace{0.5cm} { \LARGE {\bf Gaussian multi-matrix models}}

\vspace{1.1cm}

Ofer Aharony$^{\sharp}$ and Sean A. Hartnoll$^\flat$

\vspace{0.8cm}

{\it $^\sharp$ Department of Particle Physics, \\
Weizmann Institute of Science, Rehovot 76100, Israel} \vspace{0.1cm}

and
\vspace{0.1cm}

{\it  SITP, Department of Physics and SLAC, \\
Stanford University, Stanford, CA 94305, USA}
\vspace{0.4cm}

{\it $^\flat $ KITP, University of California, \\
Santa Barbara, CA 93106-4030, USA }

\vspace{0.8cm}

{\tt Ofer.Aharony@weizmann.ac.il, hartnoll@kitp.ucsb.edu} \\

\vspace{2cm}

\end{center}

\begin{abstract}
\noindent We analyze in detail a second order phase transition that
occurs in large $N$ Gaussian multi-matrix models in which the
matrices are constrained to be commuting. The phase transition
occurs as the relative masses of the matrices are varied, assuming
that there are at least four matrices in the lowest mass level. We
also discuss the phase structure of weakly coupled large $N$ $3+1$
dimensional gauge theories compactified on an $S^3$ of radius $R$.
We argue that these theories are well described at high temperatures
($T \gg 1/R$) by a Gaussian multi-matrix model, and that they do not
exhibit any phase transitions between the deconfinement scale ($T
\sim 1/R$) and the scale where perturbation theory breaks down ($T
\sim 1 / \lambda R$, where $\lambda$ is the 't~Hooft coupling).

\end{abstract}

\pagebreak
\setcounter{page}{1}

\section{Introduction}

In this paper we analyze a phase transition that occurs in large $N$
Gaussian matrix models for which the matrices commute with each
other, as a function of the masses. Our original motivation for
considering such matrix models came from the hope
\cite{Hollowood:2006xb,Hartnoll:2006pj,Gursoy:2007np} that they
would be relevant for describing the high-temperature regime of
$3+1$ dimensional weakly coupled gauge theories on $S^3$. As we
describe in section 2, it seems that this hope is not realized,
since at high temperatures the commutator squared terms in the
action are not large enough to constrain the matrices to commute
with each other. However, such matrix models have been suggested to
be useful for studying BPS states in supersymmetric gauge theories
on $S^3$ at strong coupling \cite{Berenstein:2005aa}, and we hope
that other applications may be found for them as well.

We begin in section 2 by reviewing the phase structure of weakly
coupled large $N$ gauge theories on $S^3$, and arguing that these
theories do not exhibit any phase transition in the range between
the deconfinement temperature and the temperature where perturbation
theory breaks down. In section 3 we analyze in detail a phase
transition that occurs in commuting matrix models as the masses are
varied. A special case of this transition was studied in
\cite{Gursoy:2007np}; here we generalize that analysis to general
numbers of matrices, and describe analytically how the theory looks
just below the transition. This allows us to prove that there is a
second order phase transition, and to compute the jump in the second
derivative of the action.

\section{Thermal gauge theories on $S^3$}

%\subsection{The classical action}

The thermal partition functions of gauge theories on $S^3$ can be
computed by performing the Euclidean path integral of the theory
on $S^3\times S^1$, where the $S^1$ has circumference $\beta=1/T$.
We will denote the radius of the $S^3$ by $R$.  This path integral
can naturally be performed by expanding all the fields into
Kaluza-Klein (KK) modes on $S^3\times S^1$, and then integrating
over these modes (which are just matrices).
%We use a normalization
%of the fields in our $3+1$ dimensional action in which there is a
%$1/g_{YM}^2=N/\lambda$ sitting in front of the whole action, such
%that the explicit gauge interaction terms have coefficients which
%are pure numbers.
%After the KK expansion we have an integral over
%an infinite number of matrices coming from the KK modes, with an
%action proportional to $R^3 N /\lambda T$.
The derivative terms in the original action now become mass terms
for the various matrices, with contributions proportional to $T^2$
coming from the Euclidean time derivatives, and contributions
proportional to $1/R^2$ coming from the spatial derivatives.

Classically, such a gauge theory has a single massless mode -- the
zero mode $\alpha$ of $A_0$, both on the $S^3$ and on the $S^1$. All
other modes are massive (assuming that any scalar fields either have
a classical mass, or have a conformal coupling to the curvature that
gives them a mass on $S^3$), with masses at least of order $1/R$ or
of order $T$. It is then natural to integrate out all the other
modes, and obtain an effective action for $\alpha$, which due to the
symmetry under large gauge transformations on the $S^1$ is really an
action for the unitary matrix $U = e^{i\beta \alpha}$. This
effective action turns out to be non-trivial already in the free
theory, since the one-loop path integral with some background value
for $\alpha$ is non-trivial. We assume for simplicity that all
fields in the gauge theory are in the adjoint representation; the
generalization to other cases is straightforward. The effective
action of such free gauge theories, as computed in
\cite{Sundborg:1999ue, Aharony:2003sx}, takes the form, up to an
overall additive constant,
\be \label{Uaction} S = - \sum_{n=1}^{\infty} \frac{1}{n}
\left(z_B(n \beta) + (-1)^{n+1} z_F(n \beta) \right) \Tr(U^{n})
\Tr(U^{-n}), \ee
where $z_B(\beta)$ ($z_F(\beta)$) is the generating function for the
bosonic (fermionic) modes in this classical theory on $S^3$, given
by the sum of $e^{-\beta E_i}$ over all bosonic (fermionic) states
of energy $E_i$, counting each mode in the adjoint representation
once. The phase structure of the matrix model \eqref{Uaction} was
analyzed in \cite{Sundborg:1999ue, Aharony:2003sx}; the action $S$
provides an attraction between the eigenvalues (at least at short
distance), but as in any unitary matrix model there is a repulsion
between the eigenvalues coming from the measure. We can write the
action in terms of the eigenvalues $e^{i\theta_p}$ ($p=1,\cdots,N$)
of $U$ in the form:
\be \label{thetaaction} S = \sum_{p\neq q} \sum_{n=1}^{\infty}
\frac{1}{n} \left(1 - z_B(n \beta) - (-1)^{n+1} z_F(n \beta) \right)
\cos(n(\theta_p-\theta_q)), \ee
where the first term comes from the change of measure from the
unitary matrix to its eigenvalues. From now on we will discuss only
the large $N$ limit of the gauge theory, in which we can assume that
there is some smooth distribution of the eigenvalues. At low
temperatures the repulsion wins and the eigenvalue distribution is
uniform, corresponding to a confined phase in which the expectation
values of the Polyakov-Susskind loops $\Tr(U^n)$ vanish. At a
temperature $T_{d}$ of order $1/R$, given by the solution to
$z_B(\beta)+z_F(\beta)=1$, there is a weakly first order phase
transition to a phase where the eigenvalue distribution is
non-uniform and gapped, and this deconfined phase governs the high
temperature behavior. Adding higher loop corrections turns this
weakly first order transition either into a first order transition
or into a second order transition followed by a third order
transition, depending on a coefficient which requires a three-loop
computation \cite{Aharony:2003sx, Aharony:2005bq}.

For $T \gg 1/R$ the eigenvalue distribution becomes highly
localized. In this limit the functions $z_B$ and $z_F$ go as $2 n_B
(TR)^3$ and as $2 n_F (TR)^3$, respectively, where $n_B$ ($n_F$) is
the number of bosonic (fermionic) adjoint degrees of freedom in the
theory: two from the vector field plus one for every additional
scalar field. We can then expand the action \eqref{thetaaction} for
small values of $\theta$. The action includes a quadratic term,
%Assuming
%without loss of generality that the eigenvalue distribution is
%localized near $\theta=0$, and symmetric around $\theta=0$, the
%terms in \eqref{thetaaction} coming from integrating out the other
%modes at one-loop take the approximate form
%
%\be \label{hightaction} \begin{split} S \simeq &\, 2 N \sum_{p=1}^N
%\sum_{n=1}^{\infty} \frac{1}{n} \left(-n_B
%\left(\frac{TR}{n}\right)^3 -
%(-1)^{n+1} n_F \left(\frac{TR}{n}\right)^3\right) (1 - n^2 \theta_p^2) =\\
%=&\, N (TR)^3 \left(- \frac{N \pi^4}{45} (n_B + \frac{7}{8} n_F) +
%\frac{\pi^2}{3} (n_B + \frac{1}{2} n_F) \sum_p \theta_p^2\right).
%\end{split} \ee
%
%The large temperature limit can also be thought of as a large volume
%limit, since the free theory (with no classical masses) depends only
%on the dimensionless combination $TR$. The first term in
%\eqref{hightaction} is precisely $(-\beta)$ times the free energy
%density of the free gauge theory in flat space, multiplied by the
%volume of $S^3$, as we expect to find in the large volume limit.
%
%The second term in \eqref{hightaction} looks like a one-loop mass
which is a one-loop mass term for the zero mode $\alpha$; recalling
that we are working in a normalization of the fields with a factor
of $N/\lambda$ in front of the classical action (where $\lambda$ is
the 't Hooft coupling $\lambda \equiv g_{YM}^2 N$), this quadratic
term corresponds to a physical mass proportional to $\lambda T^2
(n_B + \half n_F)$. This is precisely the one-loop ``electric mass
term'' that we expect to find for $A_0$ in the large volume limit
(see \cite{Gross:1980br} and references therein); in the large $TR$
limit the diagrams giving this mass term in the theory on $S^3$
become identical to the same diagrams on $\R^3$ (they are not
IR-divergent). The presence of this mass term means that even though
classically $\alpha$ is always the lightest mode, in the theory with
finite coupling this is no longer true when $\lambda T^2 \sim
1/R^2$, since then the KK modes of other fields could start becoming
lighter than $\alpha$, and it no longer makes sense to integrate out
the other fields and keep only $\alpha$.

In the range $TR \gg 1$ it is easy to see that the higher order
interaction terms in \eqref{thetaaction} are negligible, so the
effective action of $U$ is simply a Gaussian matrix model, and the
eigenvalue distribution approaches a semi-circle. When $T \gtrsim 1
/ \sqrt{\lambda} R$, we need to add also additional fields in our
effective action, since the lowest KK modes of the other massless
bosonic fields have masses of the same order (they have classical
masses of order $1/R$, and all of them, except for the spatial
components of the gauge field $A_i$, also have one-loop mass terms
of order $\sqrt{\lambda} T$). However, at such high temperatures,
all fields are massive enough so that their eigenvalues are small,
and all interaction terms can be ignored. Balancing the mass term
against a logarithmic eigenvalue repulsion for each field of mass
$m$, one sees that the effects of the classical commutator squared
interaction terms scale as $\lambda T / m^4 R^3$ compared to the
mass terms, so they can be consistently ignored whenever $\lambda T
\ll m^4 R^3$; this is true for all our fields until we reach $T \sim
1 / \lambda R$. At this scale perturbation theory breaks down since
the interactions of the $A_i$ fields (with $m \sim 1/R$) become
large (note that up to this temperature, perturbation theory is
valid if we add to it the effect of the ``electric mass terms'', as
discussed in section 5 of \cite{Aharony:2005bq}). The effects of
additional interaction terms arising from quantum corrections are
even smaller than the effect of these classical interactions, so
they can also be ignored for all $T \gg 1/R$.

Thus, for all $1 / \lambda R \gg T \gg 1/R$, gauge theories on $S^3$
are well-approximated by a Gaussian multi-matrix model for all the
KK modes, with the masses given by a combination of the classical KK
masses and the one-loop masses. The eigenvalue distribution of these
matrices is (to a very good approximation) a semi-circle
distribution, with a size proportional to $1/m$. There is no
correlation between the different modes, so that different modes do
not generically commute with each other\footnote{We are very
grateful to Shiraz Minwalla for pointing this fact out to us, and
correcting the first version of this paper in which we wrongly
assumed that the different modes commute.}. When $T \ll 1 /
\sqrt{\lambda} R$, $\alpha$ is much lighter than the other modes, so
its eigenvalues will be larger and dominate the dynamics. On the
other hand, when $T \gg 1 / \sqrt{\lambda} R$, the KK modes of $A_i$
will be much lighter than those of all other fields (since they are
the only ones which do not obtain an ``electric mass term'' at
one-loop order), so they will dominate the dynamics.

This implies that such weakly coupled gauge theories do not have
phase transitions in the range of temperatures between the
deconfinement scale $T \sim 1/R$ and the scale $T \sim 1 / \lambda
R$ where perturbation theory breaks down. For any $T \ll 1 /
\sqrt{\lambda} R$ the theory is well-approximated by the
single-matrix model \eqref{Uaction}, which has one or two phase
transitions around $T \sim 1/R$ but no additional transitions at
higher temperatures. For all $1/\lambda R \gg T \gg 1/R$ the
theory is well-approximated by a Gaussian multi-matrix model, and
such models do not have any phase transitions. This contradicts
previous claims about phase transitions in this range of
temperatures that were made in
\cite{Hollowood:2006xb,Gursoy:2007np}; these claims were based on
the assumption that the matrices commute with each other. This
assumption seems to be wrong (at least for $T \gg 1/R$), and the
dominant configuration, in which the matrices do not necessarily
commute, does not exhibit any phase transitions in this range.

\section{Commuting Gaussian multi-matrix models}

In this section we analyze the behavior of Gaussian multi-matrix
models of the form
\be \label{matrixaction} S = N \sum_i m_i^2
\Tr((\vec{\Phi}^i)^2),\ee
where $\vec{\Phi}^i$ is a vector of $n_i$ Hermitean matrices, and
the matrices are coupled by the requirement that they commute with
each other in the dominant configuration. Similar models have been
discussed recently in \cite{Gursoy:2007np,Berenstein:2005aa,
Berenstein:2007wz}. We will not discuss here the justification for
the assumption that the matrices commute; in general one might
expect that if there are strong enough commutator interactions that
would force the matrices to commute, they would also lead to strong
interactions between the diagonal elements, but we will assume that
imposing the constraint that the matrices commute does not lead to
any other interactions in our model. Perhaps this can be justified
in a model describing only BPS states
\cite{Berenstein:2005aa,Berenstein:2007wz}.

With the assumption that all the matrices commute, we can
diagonalize all of them simultaneously with eigenvalues
$\vec{\phi}^i_p$ ($p=1,\cdots,N$). The action for the eigenvalues,
including the term coming from the change in the measure from the
matrices to their eigenvalues, then takes the form
\be \label{oureffactionn} S = N \sum_i m_i^2 \sum_p
|\vec{\phi}^i_p|^2 -\half \sum_{p\neq q} \ln\left(\sum_i
|\vec{\phi}^i_p - \vec{\phi}^i_q|^2\right). \ee

We assume that the masses are ordered so that $m_1^2 < m_2^2 <
\cdots$; if some masses are equal we can join the corresponding
modes together into a single vector. We will discuss the phase
structure of the theory as the ratios between the masses are varied.
The equations of motion following from the action
\eqref{oureffactionn} are
\be\label{eq:eom} m_i^2 \vec{\phi}^i_p = \frac{1}{N} \sum_{q \neq
p} \frac{\vec{\phi}^i_p - \vec{\phi}^i_q}{\sum_j |\vec{\phi}^j_p -
\vec{\phi}^j_q|^2}. \ee

\subsection{The spherical phase and its stability}

In this model, unlike the general (unconstrained) Gaussian matrix
model that was mentioned in the previous section, we do not have a
separate repulsive interaction between the eigenvalues of each
matrix, but just a single repulsive potential \eqref{oureffactionn}.
Thus, the simplest assumption to make is that only the lightest mode
$\vec{\phi}^1$ condenses, and all others do not. In the large $N$
limit where we have a smooth distribution $\rho(\vec{\phi})$ for the
eigenvalues of $\vec{\phi}^1$, we can write the equation of motion
for such a solution as
\be\label{eq:eomone} m_1^2 \vec{\phi} = \int d^{n_1} \vec{\phi}'
\rho(\vec{\phi}') \frac{\vec{\phi} - \vec{\phi}'}{|\vec{\phi} -
\vec{\phi}'|^2} \,. \ee
A simple argument shows that (for $n_1 > 2$) there are no
solutions in which $\rho$ is a smooth function on $\R^{n_1}$
\cite{Berenstein:2005aa, Gursoy:2007np}. It is easy to see that \eqref{eq:eomone} has a solution
where all eigenvalues are distributed on an $S^{n_1-1}$ sphere
\cite{Berenstein:2005jq, Berenstein:2007wz, Gursoy:2007np}, of the
form
\be\label{eq:solone} \rho(\vec{\phi}) =
\frac{\d(|\vec{\phi}|-r_1)}{|\vec{\phi}|^{n_1-1} \text{Vol}
(S^{n_1-1})} \,, \ee
where the sphere has  radius
\be\label{radius} r_1 = \frac{1}{\sqrt{2} m_1}\,. \ee
The fact that the radius does not depend on the dimension was noted
in \cite{Berenstein:2005jq}. This solution is valid for all $n_1 >
1$; for $n_1=1$ the solution turns out to be a semi-circle, and for
$n_1=2$ the lowest action saddle is in fact a disc rather than the
circle described above, but in this note we will limit ourselves to
theories with $n_1 \geq 4$. The action for this configuration takes
the form
\be \label{onefaction} S = N^2 \left(\frac{1}{2} + \ln(m_1)
-\frac{1}{2} f_{n_1}\right),\ee
where $f_n$ is a complicated expression involving hypergeometric
functions, which is monotonically increasing with $n$ (approaching
zero for large $n$). Its values for small $n$ are $f_8 =
37/60-\ln(2) \simeq -0.0765$, $f_7 = -47/60+\ln(2) \simeq
-0.0902$, $f_6 = 7/12-\ln(2) \simeq -0.1098$, $f_5 = -5/6+\ln(2)
\simeq -0.1402$, $f_4 = 1/2-\ln(2) \simeq -0.1931$, $f_3 =
-1+\ln(2) \simeq -0.3069$, $f_2 = -\ln(2) \simeq -0.6931$.

We can also write down many other saddle points of
\eqref{eq:eomone}, where only some of the components of
$\vec{\phi}^1$ condense on a lower dimensional sphere of radius
\eqref{radius}. These clearly have higher action than the saddle
point above due to the monotonicity of $f_n$. It is also possible
for different sets of components to condense on different spheres,
so that the distribution is a product of expressions of the form
\eqref{eq:solone}. All possible examples for $n_1=6$ were considered
in \cite{Gursoy:2007np}, and their action was always found to be
larger than that of the $S^{n_1-1}$ distribution. We believe that
this is true for all values of $n_1 > 2$.

The stability of this saddle point with respect to fluctuations of
the $\vec{\phi}^1$ eigenvalues was checked in
\cite{Gursoy:2007np}, and it seems to be stable. There are of
course zero modes corresponding to rotations of the $S^{n_1-1}$,
but all other fluctuations seem to raise the action. Next, we can
check for the stability of this saddle point with respect to
turning on the eigenvalues of the second lightest mode,
generalizing a similar computation in \cite{Gursoy:2007np}. A
fluctuation in $\vec{\phi}^2$ leads to the quadratic action
\be \d S = N \sum_p m_2^2 |\vec{\phi}^2_p|^2 - \frac{1}{2} \sum_{p
\neq q} \frac{|\vec{\phi}^2_q - \vec{\phi}^2_p|^2}{|\vec{\phi}^1_q
- \vec{\phi}^1_p|^2} \,. \ee
In the large $N$ limit, most of these fluctuations start becoming
tachyonic when
\be\label{eq:transition}
m_2^2 < \frac{n_1-2}{n_1-3} m_1^2 \,.
\ee
This analysis requires $n_1 > 3$ for the convergence of the
relevant integral.

Thus, assuming that $n_1 > 3$, we have the following picture. If all
other masses are larger than $(n_1-2)m_1^2/(n_1-3)$, then the
dominant saddle point is the one with only $\vec{\phi}^1$ condensed
on a $S^{n_1-1}$. Then, as we change the masses (perhaps due to
changes in external parameters such as the temperature), this saddle
point becomes unstable whenever the mass of some mode goes below
this value. The crucial dynamics here is that once one set of
eigenvalues has condensed, it prevents the next lightest mode from
condensing, over a range of masses.

\subsection{The behavior just below the transition}

In order to analyze the phase transition in detail we need to know
the dominant eigenvalue distribution immediately below the
transition. This follows from the action for the joint eigenvalue
distribution of $\phi^1$ and $\phi^2$, which in the large $N$ limit
we denote as a continuous function $\rho(\phi^1,\phi^2)$. We will
perform this analysis here only for $n_1 \geq 6$. The action is
\bea\label{eq:twofields} \lefteqn{\frac{1}{N^2} S = \int
d^{n_1}\phi^1 d^{n_2}\phi^2 \rho(\phi^1, \phi^2)\left(m_1^2 |\vec
\phi^1|^2 + m_2^2 |\vec \phi^2|^2
\right)} \nonumber \\
 & & -  \frac{1}{2} \int d^{n_1}\phi^1 d^{n_2}\phi^2
d^{n_1}\phi^1{}' d^{n_2}\phi^2{}' \rho(\phi^1, \phi^2)
\rho(\phi^1{}', \phi^2{}') \ln \left(|\vec \phi^1 - \vec \phi^1{}'
|^2 + |\vec \phi^2 - \vec \phi^2{}' |^2 \right) \,. \eea
We do not know how to minimize this action in general. However,
immediately below the transition, we expect the eigenvalue
distribution to be localized so that $|\vec \phi^2| \ll |\vec
\phi^1|$. Analytic expressions for the eigenvalue distribution may
then be obtained by expanding the action (\ref{eq:twofields}) in
powers of $\phi^2$, using
\be \label{logexpand} \ln \left(|\vec \phi^1 - \vec \phi^1{}' |^2 +
|\vec \phi^2 - \vec \phi^2{}' |^2 \right) = \ln \left(|\vec \phi^1 -
\vec \phi^1{}' |^2\right) + \frac{|\vec \phi^2 - \vec \phi^2{}'
|^2}{|\vec \phi^1 - \vec \phi^1{}' |^2} + \cdots \,. \ee
There will be regions of the eigenvalue distribution where $|\vec
\phi^1 - \vec \phi^1{}'|$ will be small and this expansion is not
strictly valid there. However, our expansion is performed inside an
integral weighted by $\rho(\phi^1, \phi^2)$, so that
just below the phase transition these regions contribute negligibly
to the integral (for $n_1 > 5$).

By imposing that the $SO(n_1) \times SO(n_2)$ symmetry of the action
is not broken by the solution, the eigenvalue distribution may be
written as
\be\label{eq:ansatz} \rho(\phi^1, \phi^2) = \int  d r_2 \rho(r_2)
\frac{\d(|\vec \phi^1|-r_1(r_2)) \d(|\vec \phi^2|-r_2)}{\sqrt{1 +
(\pa r_1/\pa r_2)^2} |\vec \phi^1|^{n_1-1} |\vec \phi^2|^{n_2-1}
\text{Vol} (S^{n_1-1}) \text{Vol} (S^{n_2-1})} \,. \ee
There are two undetermined functions here, the eigenvalue density
$\rho(r_2)$ (whose integral is normalized to one) and the radius of
the $\phi^1$ sphere $r_1(r_2)$, where $r_2$ is the radius of the
$\phi^2$ sphere. When we perform a small $\phi^2$ expansion as in
\eqref{logexpand}, we should also expand the undetermined function
\be \label{eq:ellipsoid}
r_1^2(r_2) = r_1^2(0) + r_1^{(1)}(0) r_2^2 +
\cdots \,.
\ee
Our strategy is to perform this expansion in the action, and then
extremize to solve for $r_1(0)$ and $r_1^{(1)}(0)$. However,
$r_1(r_2)$ is not the only undetermined function in the ansatz
(\ref{eq:ansatz}). We must also solve for the effective eigenvalue
density $\rho(r_2)$. For the leading order behavior below the
transition it is enough to expand the action to order $r_2^4$, and
to this order (for $n_1 > 5$) the action only depends on the
eigenvalue density through the moments
\be
\langle r_2^2 \rangle \equiv \int dr_2 \rho(r_2) r_2^2 \,, \quad
\langle r_2^4 \rangle \equiv \int dr_2 \rho(r_2) r_2^4 \,.
\ee
Naively, one may  treat $\langle r_2^2 \rangle$ and $\langle r_2^4
\rangle$ as independent variables in the action. However, this
leads to inconsistent equations of motion. The reason for this is
that the Cauchy-Schwartz inequality requires that $\langle r_2^4
\rangle \geq \langle r_2^2 \rangle^2$, with equality only for the
delta function distribution, so these variables are not
independent of each other. The procedure we will use will be to
define a new variable $x$ by
\be \langle r_2^4 \rangle = x \langle r_2^2 \rangle^2 \,, \ee
and to extremize the action with respect to $\langle r_2^2
\rangle$ and $x$ subject to the constraint $x \geq 1$.

Simultaneously to the small $r_2$ expansion, we must expand about
the transition point (\ref{eq:transition}). We do this by
introducing a small parameter, $\e$, defined by
\be
m_2^2 = \frac{n_1 - 2}{n_1 - 3} m_1^2 - \e \,.
\ee
We then solve for $r_1(0), r_1^{(1)}(0)$, $\langle r_2^2 \rangle$
and $x$ in an expansion in small $\e$. Our objective is to compute
the action of the solution just below the transition to the first
non-trivial order, which is order $\e^2$.

Using the ansatz (\ref{eq:ansatz}) for the eigenvalue
distribution, one can perform the integrals over the spherical
directions in the action. Then, we can solve the equations of
motion of $r_1(0)$, $r_1^{(1)}(0)$ and $\langle r_2^2 \rangle$ in
a power series in $\e$. We find, up to the order we need:
\bea\label{eq:solution} \langle r_2^2 \rangle & = &  \frac{
n_2 (n_1-3)^2 (n_1-4) (n_1-5) \, \e}{2 m_1^4 (n_1-2) ((n_1-4) (n_1-3)^2 +
n_2 (n_1^2 - 6 n_1 + 7 + (n_1- 3)x))} + {\mathcal{O}}(\e^2)  \,, \nonumber \\
\frac{1}{r_1^2(0)} & = & 2 m_1^2 -  \frac{4 (n_1-2) m_1^4}{(n_1-4)(n_1-3)} \langle
r_2^2 \rangle
 + {\mathcal{O}}(\e^2) \,, \nonumber \\
r_1^{(1)}(0) & = & - \frac{n_1-2}{n_1-4} + {\mathcal{O}}(\e) \,.
\eea
It is interesting that the eccentricity of the ellipsoid defined
by (\ref{eq:ellipsoid}) does not tend to $1$ at the phase
transition (it does not become a pancake). As we approach the
transition from below, the eigenvalues slide along the ellipsoid,
which remains of constant shape, and accumulate at $r_2 = 0$. Note
also that the relation between $r_1$ and $r_2$ does not depend (at
leading order) on the number of modes which condense at the
transition, $n_2$. This number only enters in the solution for
$\langle r_2^2 \rangle$. The form of the ellipsoid
(\ref{eq:ellipsoid}) suggests that below the transition the
eigenvalue distribution has topology $S^{n_1+n_2-1}$. This is
supported by the numerical results in \cite{Gursoy:2007np}.

If we now try to solve the equation of motion of $x$ we see that it
has no solution. Plugging the solution \eqref{eq:solution} into the
action, we see that within the allowed range of $x \geq 1$ the
action is minimized at the boundary $x=1$, which is why this
equation cannot be satisfied. Thus, the minimal action solution is
\eqref{eq:solution} with $x=1$ (at leading order in $\e$). This
delta function behavior of the eigenvalue density to leading order
in $\e$ implies that just below the transition the eigenvalues will
be clustered around $r_2 \sim \langle r_2^2 \rangle^{1/2}$.

We can check the correctness of the results we have just stated by
comparing with numerical solutions of the equations of motion.
Figure 1 compares our results with a numerically computed eigenvalue
distribution with $N=350$ points, $n_1=6$, $n_2=1$ and $\e = 0.04$.
There is a good agreement, especially in the region where the bulk
of the eigenvalues are clustered, at $r_2 \sim \langle r_2^2
\rangle^{1/2}$. Quantitatively, the numerical analysis gives
$\langle r_2^2 \rangle \approx 0.0073$ whereas our formula
(\ref{eq:solution}) gives $\langle r_2^2 \rangle \approx 0.0053$.
The discrepancy here is of order $\e^2 = 0.0016$, as we should
expect. The numerics furthermore give $\langle r_2^4 \rangle/\langle
r_2^2 \rangle^2 \approx 1.21$, which is roughly of order $\e$ away
from $x=1$, again consistent with our approximations.

\begin{figure}[h]
\begin{center}
\epsfig{file=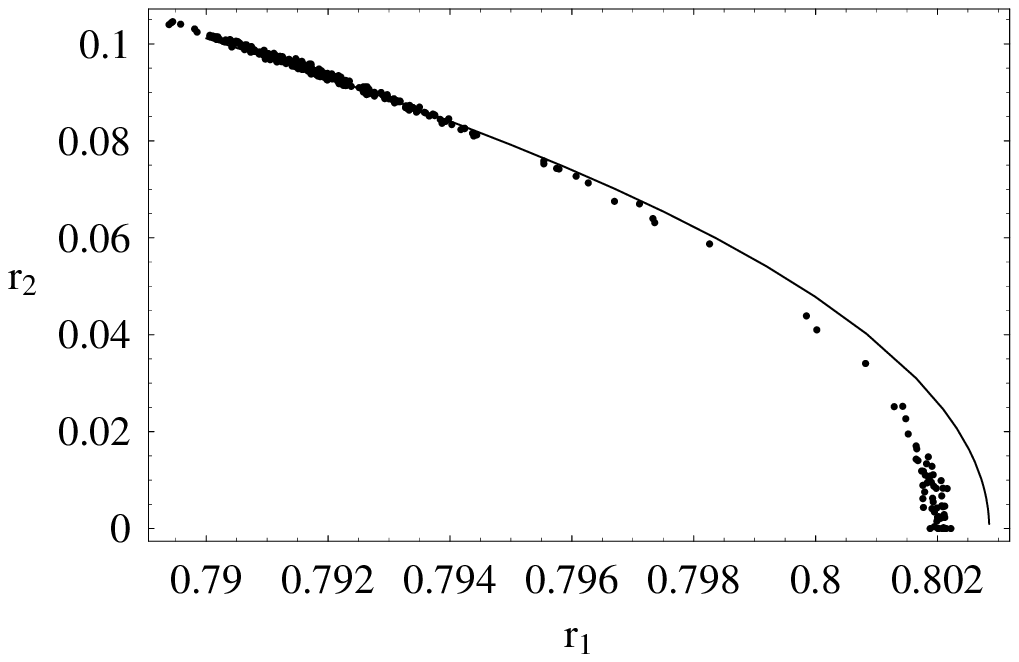,height=6cm}
\end{center}

\noindent {\bf Figure 1:} Numerical analysis and leading order
analytic results for $n_1=6, n_2=1$ at masses $m_1^2 = 0.78$ and
$m_2^2=1$, corresponding to $\e = 0.04$. The numerical analysis uses
$N = 350$ eigenvalues. At leading order in $\e$ the distribution is
supposed to be a delta function at $r_2 = \sqrt{\langle r_2^2
\rangle} = 0.0727$. Consistently with this, the fraction of the
eigenvalues in the lower grouping is around $15$ percent, which is
of order $\e$.
\end{figure}

It is now straightforward to take our solution (\ref{eq:solution})
and evaluate the action (\ref{eq:twofields}) to order $\e^2$.
Relative to the sphere solution (\ref{eq:solone}), the action is
\be \label{actiondiff} \frac{1}{N^2} \Delta S = -\frac{\e}{2}
\langle r_2^2 \rangle = - \frac{n_2 (n_1 - 5) (n_1-3)^2}{4 (n_1 - 2)
((n_1-3)^2 + n_2 (n_1-1))} \frac{\e^2}{m_1^4}\,. \ee
Thus we see that there is a second order phase transition as $\e
\to 0$.

It would be interesting to see if there are additional transitions
in these models, for instance as the mass of a third matrix is
decreased. It would also be interesting to generalize our analysis
to matrix models that have lower values of $n_1$. For $n_1 < 4$ it
seems that more than one vector of matrices condenses for any ratio
of masses, and it would be interesting to see if similar phase
transitions to the one we described still occur or not.

\section*{Acknowledgements}

We are very grateful to S. Minwalla for pointing out a mistaken
assumption in the original version of this paper and for very
helpful discussions. We would like to thank S. P. Kumar, S. Shenker,
M. Unsal and M. Van Raamsdonk for useful discussions. The work of OA
is supported in part by the Israel-U.S. Binational Science
Foundation, by a center of excellence supported by the Israel
Science Foundation (grant number 1468/06), by the European network
HPRN-CT-2000-00122, by a grant from the G.I.F., the German-Israeli
Foundation for Scientific Research and Development, and by a grant
of DIP (H.52). The work of SH was supported in part by the National
Science Foundation under Grant No. PHY05-51164.

\end{document}